\documentclass[
amsmath,amssymb,
aps
pra,
onecolumn,
superscriptaddress
]{revtex4-2}
\usepackage[utf8]{inputenc}
\usepackage{amsmath}
\usepackage{graphicx}
\usepackage{appendix}
\usepackage{color}
\usepackage{hyperref} 
\usepackage{comment}
\usepackage{tikz}
\usetikzlibrary{arrows.meta}

\begin{document}
\title{Universality of order statistics for Brownian reshuffling}
\author{Zdzislaw Burda}\email{zdzislaw.burda@agh.edu.pl} 
\affiliation{\href{https://ror.org/00bas1c41}{AGH University},
Faculty of Physics and Applied Computer Science,
al. Mickiewicza 30, 30-059 Krak\'ow, Poland}

\author{Mario Kieburg}\email{m.kieburg@unimelb.edu.au}
\affiliation{\href{https://ror.org/01ej9dk98}{University of Melbourne}, School of Mathematics and Statistics, 813 Swanston Street,
Parkville, Melbourne VIC 3010, Australia;}

\author{Tomasz Maciocha}\email{tomasz.maciocha@agh.edu.pl} 
\affiliation{\href{https://ror.org/00bas1c41}{AGH University},
Faculty of Physics and Applied Computer Science,
al. Mickiewicza 30, 30-059 Krak\'ow, Poland}

\begin{abstract}
We discuss the order statistics of the particle positions of a gas of $N$ identical 
independent particles performing Brownian motion in one dimension in a potential 
that asymptotically behaves like $V(x) \sim x^\gamma$ for $x\rightarrow+\infty$, with a positive
power $\gamma>0$. We show that in the stationary state, the order statistics 
that describe how the leaders are reshuffled are universal and independent
of $\gamma$. What depends on $\gamma$ is the timescale of the leaders' reshuffling, which scales as a power of the logarithm of the
population size: $t \sim (\ln N)^\frac{2(1-\gamma)}{\gamma} \tau$, where $\tau$ is of order one. 
We derive the probability that the particle which has the $k$th largest value of $x$ at some time
$t_1$ will have the $j$th largest value at time $t_2=t_1+t$ 
in the form of an explicit expression for the generating function for the reshuffling 
probabilities for all $k\ge 1$ and $j\ge 1$. 
The generating function, expressed in scaled time $\tau$, 
is independent of $\gamma$. In particular, we show that the average percentage overlap 
coefficient of leader lists takes the universal, $\gamma$-independent 
form ${\rm erfc}(\sqrt{\tau})$ for long lists.
\end{abstract}

\maketitle

\section{Introduction \label{sec:intro}}

Rankings and order statistics are important in data science, statistics, and statistical physics. 
Rankings are related to extreme values, so-called outliers — elements of the population that lie at the edge of the population or beyond it.
The study of extreme values is a well-established field of mathematical statistics and statistical physics~\cite{ranking,os1,os2,os3,MPS} and still attracts a lot of interest; see~\cite{KLD,HCCC,HCC,DDP,LX}. 
If one considers important traits and orders the members of a population according to them, the leaders, high-ranking members, and outliers are usually those
responsible for the direction in which the population's evolution will take place~\cite{O,JK,BsM,JRBO,AMS}. Therefore, while discussing the evolution 
of a complex system, one may be interested in questions concerning ranking reshuffling, which will provide insight into what happens
on the edge of the population; for example, how long the leader will remain the leader or what 
the chances are of finding $m$ out of $n$ current leaders on the top-$n$ list at some later time, {\em etc}. 
Recently, such dynamical aspects of rankings have attracted attention \cite{LeD,MS,BK,BGFSBBB,BKMS,IPGB,WK,DCLMN,KM,DHJX,DDMS2017,EM2008}.

The answers to these questions obviously depend on the details of the dynamics. In the present work, we consider random dynamics 
driven by independent and identical random changes in the quantity being ranked. 
A population subjected to such random dynamics can be modeled as a one-dimensional 
gas of particles which, in the simplest setting, performs independent and identical Brownian motion in a potential. 
In this model, the position of a particle on the real axis simulates the quantity being ranked, with the rightmost particle, {\em i.e.}, 
the one with the highest coordinate value on the real axis, being assumed to have rank one and the next particle having 
rank two, {\em etc.}. A simple quantity measuring this reshuffling rate is the overlap ratio $\Omega_n(t)$, which measures the fraction of $n$ leaders that are still among $n$ leaders after time $t$. The overlap ratio is a particular example of the Szymkiewicz–Simpson coefficient used as a similarity
measure of two sets~\cite{stat-overlap}. In Ref.~\cite{BK}, we have derived an analytical formula for the expectation value of the overlap ratio for a linearly growing potential with a hard wall at the origin. The formula for the overlap drastically simplifies for large $n$, which is much smaller than the population size: $1 \ll n\ll N$. The limit law has the simple form
\begin{equation}
\langle \widehat{\Omega}_\infty(\tau)\rangle \approx \mbox{ erfc}(\sqrt{\tau})
\label{eq:omega_inf}
\end{equation}
after expressing it in the appropriately rescaled time $\tau$.

In the present work, we show that the random
reshuffling dynamics leads, in the limit of infinite population size $N\to\infty$, to the very same limit regardless of the details of the potential, which we assume is confining and behaves like $V(x)\propto x^\gamma$ with $\gamma>0$ for $x\to+\infty$. Therefore, the law~\eqref{eq:omega_inf} is universal for this class of stochastic processes. It can therefore be treated as a benchmark for rank reshuffling.

This article is organized as follows. In Section~\ref{sec:prelim}, 
we recall the basic equations describing the diffusion of a gas of $N$ particles in a potential in one dimension to prepare the ground and introduce our notation.
In Section~\ref{sec:resh_prob}, we define the reshuffling probabilities, which are the probability that a particle
having rank $j$ at time $t_1$ will have rank $k$ at time $t_2$. Furthermore, we introduce
a generating function for the reshuffling probabilities. It is the main object 
of the analysis from which one can derive all important quantities describing order statistics. 
In this section, we also define the overlap ratio between the top-$n$ ranking lists.
In Section~\ref{sec:scaling}, we discuss the scaling of the position of the leader and the scaling of the typical time for their reshuffling.
We show how to introduce scaling parameters in which one can conveniently
express the asymptotic behavior of the reshuffling probabilities and
the order statistics. In Section~\ref{sec:const_drift}, we recall the main steps of the
derivation of the generating function for diffusion with a constant negative drift and a reflective wall~\cite{BK} which highlights the main ideas and summarizes the main results.
These methods will be applied to the Ornstein-Uhlenbeck process in Section~\ref{sec:OU}, which is diffusion in a quadratic potential. 
We identify the scaling relations and show that, after proper rescaling of the parameters,
we obtain identical results to those for the diffusion with a constant negative drift and a reflective wall, which has been numerically observed in~\cite{BK}. We generalize this argument to potentials that behave asymptotically as $V(x)\propto x^\gamma$ 
for $x\rightarrow +\infty$ with $\gamma>0$, in Section~\ref{sec:univ}. This shows the universality of these results. To highlight that those reshuffling 
laws can be found even in nonstationary stochastic processes, we studied free diffusion with normal initial conditions in Section~\ref{sec:free}.
We show that rank reshuffles in free diffusion can be essentially mapped to rank reshuffles in the Ornstein-Uhlenbeck process.
In section ~\ref{sec:concl}, we summarize and conclude this work.

\section{1-dim Brownian Motion in a Potential \label{sec:prelim}}

We consider a gas of $N$ independent particles performing Brownian motion
in a potential $V(x)$. The force resulting from this potential is $f(x)=-V'(x)$. It plays the role of a position-dependent drift. 
The Brownian motion of particles in this potential is described by $N$ identical equations
\begin{equation}
dx_i= -V'(x_i) dt + \sigma dB_i(t), \quad i=1,\ldots,N,
\label{eq:ito}
\end{equation}
where $x_i$ are the positions of the particles at 
time $t$, and $B_i(t)$ are independent centered Wiener processes with the trivial covariance structure
$\langle dB_i(t) dB_j(t) \rangle = \delta_{ij} dt$. The corresponding Fokker-Planck equation for 
the probability density $p(x,t)$ describing the distribution
of particle positions at time $t$ reads
\begin{equation}
    \partial_t p(x,t) = \partial_x [V'(x) p(x,t)] + \frac{\sigma^2}{2} \partial_{xx}^2 p(x,t) .
    \label{eq:FP}
\end{equation}
To go from \eqref{eq:ito} to \eqref{eq:FP}, the It\^o calculus was used.
The probability density at time $t_2=t_1+t$ is related to that at an earlier time $t_1$ by a propagator $W$
\begin{equation}
    p(x,t_2) = \int_{-\infty}^{+\infty} dy p(y,t_1) W(y,x,t_2-t_1) .
\end{equation}
The propagator $W$ is also called the Green function or the heat kernel. The propagator depends on the time difference $t=t_2-t_1$ because
the diffusion equation~\eqref{eq:FP} is time-translation invariant. The Green function satisfies the following initial value problem
\begin{equation}
    \partial_{t} W(y,x,t) = \partial_x [V'(x) W(y,x,t)] + \frac{\sigma^2}{2} \partial_{xx}^2 W(y,x,t) 
    \label{eq:W}
\end{equation}
and
\begin{equation}
\lim_{t\rightarrow 0^+} W(y,x,t)=\delta(x-y) .
\end{equation}

For a wide class of potentials, the system has a unique stationary state, given by
\begin{equation}
  p(x) = c e^{-2V(x)/\sigma^2},
  \label{eq:pV}
\end{equation}
defined for all $x$ in the support of the function $V(x)$. We assume that the support of $V(x)$
is connected. The system has a unique stationary state if the potential falls off to zero quickly enough for $x\rightarrow \pm \infty$; more
precisely, if there is an $\epsilon>0$ such that $|x|^{1+\epsilon} e^{-2V(x)/\sigma^2} \rightarrow 0$ for $x\rightarrow \pm \infty$. 
The constant $c$ guaranties the probabilistic normalization $\int p(x) dx =1$. The probability density function of the 
stationary state is related to the heat kernel as follows
\begin{equation}
    \lim_{t\rightarrow \infty} W(y,x,t) = p(x) 
\end{equation}
and satisfies the stability property
\begin{equation}
    p(x) = \int_{-\infty}^{+\infty} dy p(y) W(y,x,t) .
\end{equation}
To avoid introducing new symbols, we denoted the probability density function of the stationary state  by the same letter $p$ as for the probability density function of the distribution 
at time $t$. This does not lead to confusion because, in the first case, the 
function has one argument, and in the second case, two. In particular, in the stationary state, we simply set $p(x,t)=p(x)$.

At any time, the particles can be ordered with respect to their positions 
$x_1\ge x_2 \ge \ldots \ge x_N$, where strict inequality holds almost surely, meaning the event that two particles occupy the same position at a specific time $t>0$ has a vanishing probability. We are interested in how the ordering (ranking of the particles) 
changes under diffusion in a potential that, for $x\rightarrow+\infty$, 
behaves asymptotically as
\begin{equation}
  V(x) \sim \frac{k}{2} x^\gamma
  \label{eq:poten}
\end{equation}
where $k$ is a positive constant and $\gamma>0$. More precisely, we are interested in questions
like: what is the probability that a particle with rank $k$ at time $t_1$ will have rank $j$ at 
a later time $t_2=t_1+t$? Another question is what the probability is that
the leader at $t_1$ will still be the leader at time $t_2$. Or, how many of the $n$ leaders at $t_1$ will still be among the $n$ leaders at $t_2$? 

As we shall see, all that matters 
in the limit $N\rightarrow \infty$ is the asymptotic behavior of the potential. All other details do not matter for the statistics of the leaders' rankings. 
However, to be specific, we can think of diffusion in a symmetric potential
$V(x)=\frac{1}2 k |x|^\gamma$ or in a potential $V(x) = \frac{1}{2} k x^\gamma$ 
for $x\ge 0$ that has a reflective wall at $x=0$.

For the sake of simplicity and convenience, we will choose units of time and length such that, in these
units, $\sigma=k=1$. Any function expressed in these units can be converted to units in which 
$\sigma$ and $k$ have arbitrary values by the transformation
\begin{equation}
\label{eq:transf}
    x =  \left(\frac{\sigma^2}{k}\right)^{1/\gamma} x_*
    \qquad{\rm and}\qquad
    t = \frac{1}{\sigma^2} \left(\frac{\sigma^2}{k}\right)^{1/\gamma} t_* 
\end{equation}
where $t_*$ and $x_*$ correspond to the units in which $\sigma=k=1$. 
Therefore, without loss of generality, we set $\sigma=k=1$ from here on.

\section{Reshuffling probabilities \label{sec:resh_prob}}

To describe the ranking statistics, let us define the reshuffling probability
$p_R(k,j,t)$ as the probability that a particle with rank $k$ at time $t_1$ will have rank $j$
at time $t_2=t_1+t$. In a stationary state, the probability depends only on the time difference $t=t_2-t_1$.

The reshuffling probabilities can be expressed in terms of the complementary cumulative distribution function 
\begin{equation}
  P_+(x) = \int^{+\infty}_x d u p(u).
\end{equation}
The cumulative distribution function is $\int_{-\infty}^x d u p(u)=1-P_+(x)$. 
Since we are mainly interested in stationary states, we skipped the dependence on time.

Similarly, we can define cumulative distribution functions for a two-point probability distribution
\begin{equation}
  q(y,x,t) = p(y) W(y,x,t)=q(x,y,t),
  \label{eq:q}
\end{equation}
which describes the probability that a given particle is at $y$ at time $t_1$ and at $x$ at time $t_2=t_1+t$. The marginal distribution
of $q$ is equal to $p$: $\int dx q(y,x,t) = \int dx q(x,y,t) = p(y)$.
For a two-point probability, there are four possible cumulative distributions $Q_{\pm \pm}(y,x,t)$
that correspond to the probabilities of finding a particle  above or below $y$ at time $t_1$ and above or below $x$ at time $t_2$. 
The probability of events above the thresholds $x$ and $y$ is
\begin{equation}
    Q_{++}(y,x,t) = \int^{+\infty}_y \int^{+\infty}_{x} q(v,u,t) dv d u .
    \label{eq:Qpp_q}
\end{equation}
If we replace the limits in the first integral with $(-\infty,y)$ or in the second integral with $(-\infty,x)$, we obtain the other probabilities given by
the relations
\begin{equation}
\begin{split}
Q_{-+}(y,x,t)& =P_+(x)-Q_{++}(y,x,t),\\ Q_{+-}(y,x,t)& =P_+(y)-Q_{++}(y,x,t),\\
Q_{--}(y,x,t)& = 1 - P_+(y) - P_+(x) + Q_{++}(y,x,t),
\end{split}
\label{eq:srules}
\end{equation}
which follow from the marginal distributions.

Using these probabilities, we can express ranking statistics in an elegant way. For example, the probability that a leading particle, 
having rank $1$ at time $t_1$, will have rank $1$ at a later time $t_2=t_1+t$ is given by
\begin{equation}
    p_R(1,1,t)=N \int_{-\infty}^{+\infty} \int_{-\infty}^{+\infty} dy dx q(y,x,t) \left(Q_{--}(y,x,t)\right)^{N-1} .
    \label{eq:p11}
\end{equation}
The interpretation of the equation is as follows. Consider a reference particle that is at position $y$ at time $t_1$ and at position $x$ at time $t_2=t_1+t$.
Due to the independence of the particle positions, the factor $\left(Q_{--}(y,x,t)\right)^{N-1}$ is the probability
that all other particles are at positions smaller than $y$ at time $t_1$ and smaller than $x$ at time $t_2$. 
This reflects the fact that the reference particle is indeed the leader at both $t_1$ and $t_2$. The integral sums the contributions from all possible positions $y$ and $x$ according to the probability measure $q(y,x,t)dy dx$.
Finally, the factor $N$ in front of the integral comes from $N$ different leaders because the leader can be any of the $N$ particles.  

To generalize~\eqref{eq:p11} to an arbitrary reshuffling probability $p_R(k,j,t)$ for any $j,k=1,\ldots,N$, 
it is convenient to introduce a generating function
\begin{equation}
    P_R(z,w,t) = \sum_{k=1}^\infty \sum_{j=1}^\infty p_R(k,j,t) z^{k-1} w^{j-1} .
    \label{eq:PRdef}
\end{equation}
It can be argued that the generating function must have the following form,
\begin{equation}
    P_R(z,w,t) = N \int_{-\infty}^{+\infty} \int_{-\infty}^{+\infty} dy dx q(y,x,t) \left[Q_{--}(y,x,t) + zQ_{+-}(y,x,t) + wQ_{-+}(y,x,t) + zw Q_{++}(y,x,t)\right]^{N-1} .
     \label{eq:PRrel}
\end{equation}
The argument is as follows. 
The reshuffling probabilities $p_R(k,j,t)$ can be identified from the expansion of the last expression
\begin{equation}
    P_R(z,w,t) = N \sum_{\{a,b,c,d\}} z^{b+d} w^{c+d} \frac{(N-1)!}{a!b!c!d!} \int_{-\infty}^{+\infty} dy \int_{-\infty}^{+\infty} 
    dx q(y,x,t) Q^{a}_{--}(y,x,t)Q^b_{+-}(y,x,t) Q^c_{-+}(y,x,t) Q^d_{++}(y,x,t) .
    \label{eq:PRexp}
\end{equation}
The sum is over all non-negative integers $a,b,c,d$ satisfying $a+b+c+d=N-1$. In the integrand, $y$ is the position of the reference particle at time $t_1$, and $x$ is the position of the reference
particle at time $t_2=t_1+t$. The index 
$a$ refers to the number of particles that are below positions $x$ and $y$. The index $b$ refers to the number of particles below $x$ and above $y$. $c$ is the number of particles above $x$ and below $y$, and $d$ is the number of particles that are both above the positions $x$ and $y$. 
The sum over all coefficients at the power $k-1=b+d$ of $z$ and the power $j-1=c+d$ of $w$ in the series expansion~\eqref{eq:PRexp} 
gives the probabilities $p_R(k,j,t)$ that the reference particle has rank $k$ at $t_1$. This is because there are $k-1$ particles
above it at $t_1$, and that it has rank $j$ at $t_2$ since only $j-1$ particles are above it at $t_2$. 
The multinomial coefficient $(N-1)!/(a!b!c!d!)$ counts all combinations of placing $(N-1)$ particles in the four segments
$\pm,\pm$. The $N$-th particle is represented by $q(y,x,t)$ and the factor $N$ since it can be any of the $N$ particles.

Using the sum rules \eqref{eq:srules}, we can write the generating function~\eqref{eq:PRrel} in terms of the complementary cumulative probabilities,
\begin{equation}
    P_R(z,w,t) = N \int_{-\infty}^{+\infty} \int_{-\infty}^{+\infty} dy dx q(y,x,t) 
    \left[1 - (1-z)P_+(y) - (1-w) P_+(x) + (1-z)(1-w) Q_{++}(y,x,t) \right]^{N-1} 
     \label{eq:PRrel_extreme}
\end{equation}
This form will be particularly useful in the analysis of extreme value statistics, see the next section.

From the generating function~\eqref{eq:PRdef}, we can easily calculate the reshuffling probabilities via Taylor coefficients
\begin{equation}
    p_R(k,j,t) = \frac{1}{(k-1)!}\frac{\partial^{k-1}}{\partial z^{k-1}} \frac{1}{(j-1)!}\frac{\partial^{j-1}}{\partial w^{j-1}} P_R(z,w,t)\bigg|_{z=w=0} .
    \label{eq:pRpartial}
\end{equation}
An alternative way is to treat the arguments of the generating function $z$ and $w$ as complex variables and to use the residue theorem
\begin{equation}
    p_R(k,j,t) = \oint \frac{dz}{2\pi i z^k} \oint \frac{dw}{2\pi i w^j} P_R(z,w,t), 
    \label{eq:pRres}
\end{equation}
where the integrals are over small contours around the origin, especially $|z|=|w|=\epsilon< 1$.
Note that for $k=j=1$, we immediately recover \eqref{eq:p11}.

In addition to the reshuffling probabilities, we can 
introduce another quantity that we call the overlap ratio or overlap coefficient, which, in a very informative and intuitive way, 
captures information about the rate of reshuffling of the leaders. We consider the ranking lists of $n$ leaders at times $t_1$ and $t_2$ and denote by $n_*(t_1,t_2)$ the number of leaders appearing on both lists at times $t_1$ and $t_2$.
The overlap coefficient, also known as the overlap ratio, is in this case defined as the quotient
\begin{equation}
\Omega_n(t_1,t_2) = \frac{n_*(t_1,t_2)}{n}
\end{equation}
which determines the fraction of how many of the $n$ leaders at time $t_1$ will be among the $n$ leaders at time $t_2$.

In the steady state, the average overlap coefficient $\langle \Omega_n(t_1,t_2) \rangle$ only depends on the 
time difference $t=t_2-t_1$, so we will write it as $\langle \Omega_n(t) \rangle$. This quantity is related
to the reshuffling probabilities by the sum
\begin{equation}
\langle \Omega_n(t) \rangle = \frac{1}{n}  \sum_{k=1}^n \sum_{j=1}^n p_R(k,j,t) .
\end{equation}
Using~\eqref{eq:pRres}, we find
\begin{equation}
\langle \Omega_n(t) \rangle = \frac{1}{n} \oint \frac{dz}{2\pi i (1-z) z^n} \oint \frac{dw}{2\pi i (1-w) w^n} P_R(z,w,t) 
\label{eq:Omegan}
\end{equation}
when employing the geometric sum and using $\oint \frac{dz}{2\pi i (1-z) }  P_R(z,w,t)=\oint \frac{dw}{2\pi i (1-w) }  P_R(z,w,t)=0$. These integrals
are equal to zero because no singularity is encircled by the integration contours. 
The last equation for $\langle \Omega_n(t) \rangle$ can be alternatively written as 
\begin{equation}
\langle \Omega_n(t) \rangle = \frac{1}{n} \oint \frac{dz}{2\pi i z^n} \oint \frac{dw}{2\pi i  w^n} Z(z,w,t)
\label{eq:OmegaZ}
\end{equation}
with the function
\begin{equation}
    Z(z,w,t) = \frac{P_R(z,w,t)}{(1-z)(1-w)} = \sum_{n=1}^\infty \sum_{m=1}^\infty O_R(n,m,t) z^n w^m
    \label{eq:Z}
\end{equation}
and the coefficients
\begin{equation}
    O_R(n,m,t) = \sum_{k=1}^n \sum_{j=1}^m p_R(k,j,t) . 
\end{equation}
The coefficients $O_R(n,m,t)$ in the series expansion \eqref{eq:Z} are equal to the average number 
of particles that are in the top-$n$ list at time $t_1$ and the top-$m$ list at time $t_2=t_1+t$.
The quantity $O_R(n,m,t)$ can be interpreted as the average overlap of the two lists. This means that $Z(z,w,t)$ is a generating 
function for average overlaps. If we divide the average overlap by $\min(n,m)$, 
we obtain the overlap ratio. We are mainly interested in the case $m=n$, meaning $\langle\Omega_n(t)\rangle=O_R(n,n,t)/n$.

\section{Scaling of leaders' positions and of reshuffling time \label{sec:scaling}}

In this section, we aim to find rescaled coordinates of particle positions that focus on the range where leaders are located 
and to determine the proper time scale for leaders reshuffling. As we will see, in general, the rescaled coordinates and the rescaled time
will depend on the population size.

Consider the population of size $N$. The leader at time $t_1$ is located at $y_{max,N} = \max(y_1,\ldots,y_N)$,
and at time $t_2$ at $x_{max,N} = \max(x_1,\ldots,x_N)$. The probability distribution for the position of the leader is
\begin{equation}
    {\rm Prob}(y_{max,N}<y) = (1 - P_+(y))^N \overset{N\gg1}{\approx} \exp(- N P_+(y)) .
    \label{eq:ev}
\end{equation}
and similarly for $t_2$: ${\rm Prob}(x_{max,N}<x) {\approx} \exp( -N P_+(x))$. The idea is to introduce rescaled variables $\xi$ and $\zeta$ 
\begin{equation}
\begin{split} 
    y= a_N + b_N \xi \qquad{\rm and}\qquad
    x= a_N + b_N \zeta,
    \label{eq:scaling.position}
\end{split}
\end{equation}
with coefficients $a_N, b_N$ that depend on the population size $N$, 
so that the probability distribution for the position of leaders becomes asymptotically independent of $N$ for $N\rightarrow \infty$.
This goal is achieved if we find $a_N$ and $b_N$ such that
\begin{equation}
\begin{split} 
    P_+(y) & =  \frac{1}{N} \widehat{P}_+(\xi) + o\left(N^{-1}\right)\ ,\qquad 
    P_+(x)  =  \frac{1}{N} \widehat{P}_+(\zeta) + o\left(N^{-1}\right)
\end{split}
   \label{eq:sf.P}
\end{equation}
where $\widehat{P}_+(\xi)$ is a function that depends only on the rescaled coordinate and is independent of $N$.
The map ~\eqref{eq:scaling.position} is a standard relation in extreme value statistics used to analyze the local scale where the leaders are located. We also need a similar scaling relation for time to capture the timescale of leaders' reshuffling.
This timescale is determined from the probability $Q_{++}(y,x,t)$. For large $N$, 
$Q_{++}(y,x,t)$ is at most of order $1/N$, when $P_+(x)$ and $P_+(y)$ are of order $1/N$, because of the bound
\begin{equation}
    Q_{++}(y,x,t)\leq \min\{P_+(x), P_+(y)\}\approx\frac{1}{N}\min\{\widehat P_+(\zeta), \widehat P_+(\xi)\} .
\end{equation}
The correct timescale
\begin{equation}
\begin{split} 
    t= c_N \tau,
    \label{eq:scaling.time}
\end{split}
\end{equation}
is when $Q_{++}(y,x,t)$ is also of order $1/N$. Choosing the parameters $\tau$ \eqref{eq:scaling.time} and $\xi,\zeta$ \eqref{eq:scaling.position}, 
we obtain
\begin{equation}
\begin{split} 
    q(y,x,t) dy dx & =   \left(\frac{1}{N}\widehat{q}(\xi,\zeta,\tau)  +o\left(N^{-1}\right)\right)d\xi d\zeta,\qquad
    Q_{++}(y,x,t) =  \frac{1}{N} \widehat{Q}_{++}(\xi,\zeta,\tau) + o\left(N^{-1}\right) .
\end{split}
   \label{eq:sf}
\end{equation}
Note that on a timescale where $Q_{++}(y,x,t)$ would fall to zero faster than $1/N$ for $N\rightarrow\infty$, the term $Q_{++}(y,x,t)$ would disappear from the generating function \eqref{eq:PRrel_extreme} in the limit $N\rightarrow\infty$.
This would correspond to completely independent, randomly reshuffled populations at times $t_1$ and $t_2=t_1+t$, which means that the mixing of the leader with the rest of the population would occur instantaneously for large $N$. In other words, the most interesting timescale is the one where $Q_{++}(y,x,t)$ scales like $1/N$ for large $N$ \eqref{eq:sf}.

In summary, by changing the variables $(y,x,t)$ to $(\xi,\zeta,\tau)$, which are of order $1$,~\eqref{eq:scaling.position} and~\eqref{eq:scaling.time},
we can see how the ranking slowly changes in the time parameter $\tau$. 
As we will see, the reshuffling probabilities as functions of the scaled parameters $(\xi,\zeta,\tau)$ exhibit universal behavior for $N\rightarrow \infty$.
The generating function~\eqref{eq:PRrel_extreme} in the new rescaled time~\eqref{eq:scaling.time} will be denoted by
\begin{equation}
    \widehat{P}_R(z,w,\tau) = \lim_{N\rightarrow \infty} P_R(z,w,t= c_N \tau) .
\end{equation}
It is equal to \eqref{eq:PRrel_extreme}
\begin{equation}
    \widehat{P}_R(z,w,\tau) = \int_{-\infty}^{+\infty} \int_{-\infty}^{+\infty} d\xi d\zeta \widehat{q}(\xi,\zeta,\tau) 
    \exp\left( - (1-z)\widehat{P}_+(\xi) - (1-w) \widehat{P}_+(\zeta) + (1-z)(1-w) \widehat{Q}_{++}(\xi,\zeta,\tau) \right) .
     \label{eq:PRmain}
\end{equation}
As we will see, a natural way of calculating this integral is to change the integration variables 
to the center of mass $\alpha=(\xi+\zeta)/2$ and the relative position $\beta=\xi-\zeta$ of the leaders at times $t_1$ and $t_2$. 
The time $t=c_N \tau$, where $\tau$ is of order one, provides a time scale suitable for examining changes 
in the relative positions of the leaders $\beta=\xi-\zeta$.

As we will show in the following sections, the generating function \eqref{eq:PRmain} takes a universal form in the limit $N\rightarrow \infty$, which is independent of the power $\gamma$ governing large $x$ asymptotics of the potential $V(x)$ \eqref{eq:poten}, while the way the coefficients $a_N$, $b_N$, $c_N$ scale with the population size $N$ depends explicitly on $\gamma$. From this scaling, one can deduce how the 
leader positions and leader shuffling time depend on $N$.

\section{Diffusion with constant drift and reflective wall \label{sec:const_drift}}

In this section, we discuss diffusion with a constant drift, $\gamma=1$ \eqref{eq:poten}, in the presence of a reflective wall.
The drift is toward the wall, which causes the system to have a stationary state.
We assume that the wall is placed at $x=0$, and diffusion takes place on the real positive half-axis. 

The problem was discussed in detail in~\cite{BK}. To provide a reference point for calculations for other potentials, 
we summarize the main calculation steps here.
The Fokker-Planck equation \eqref{eq:FP} for $k=1$ and $\sigma=1$ reads
\begin{equation}\label{FP.eq.constant.drift}
    \partial_t p(x,t) = \frac{1}{2} \partial_x p(x,t) + \frac{1}{2} \partial_{xx}^2 p(x,t).
\end{equation}
The equation is supplemented with the reflecting wall condition
\begin{equation}
p(0,t) + \partial_x p(0,t) = 0
\end{equation}
which holds at any time $t$ during the evolution and ensures
that the probability is in the range $[0,+\infty)$. The potential corresponding to this setting is
\begin{equation}
    V(x)=\begin{cases}
        x/2, & x>0,\\ +\infty, & x\leq0
    \end{cases}
\end{equation}
with a constant linear growth on the positive real line, {\em i.e.}, $\gamma=1$.

The propagator is relatively complex \cite{BK,AW} 
\begin{equation}
    W(y,x,t) = \frac{e^{-\frac{(2 (y-x)-t)^2}{8 t}}}{\sqrt{2\pi t}}
 + \frac{e^{-x} e^{-\frac{(2 (x + y)-t)^2}{8 t}}}{\sqrt{2\pi t}} +
 \frac{e^{-x}}{2}  {\rm erfc}\left(\frac{2 (x + y)-t}{\sqrt{8t}}\right) 
 \label{eq:Wd}
\end{equation}
for $x,y\ge 0$, while the probability density function and the cumulative distribution function of the stationary state are extremely simple,
\begin{equation}
    p(x) = P_+(x) = e^{-x} .
\end{equation}
It can be readily checked that the two-point function $q(y,x,t) = e^{-y} W(y,x,t)$ is symmetric $q(y,x,t)=q(x,y,t)$.
The cumulative distribution for the two-point function~\eqref{eq:Qpp_q} takes the form
\begin{equation}
\begin{split}
Q_{++}(y,x,t)= & \frac{1}{2} e^{-y} {\rm erfc}\left(\frac{t+2(x-y)}{\sqrt{8t}}\right) + \frac{1}{2} e^{-x} {\rm erfc}\left(\frac{t-2(x-y)}{\sqrt{8t}}\right) \\
+ &\frac{1}{2} e^{-x-y} {\rm erfc}\left(\frac{-t+2(x+y)}{\sqrt{8t}}\right) + \frac{1}{2} {\rm erfc}\left(\frac{t+2(x+y)}{\sqrt{8t}}\right) .
\end{split}
\label{eq:qpp_drift}
\end{equation}

Following the ideas of Section~\ref{sec:scaling}, we need to find the proper scales $a_N$, $b_N$, and $c_N$; see~\eqref{eq:scaling.position} and~\eqref{eq:scaling.time} that lead to the $1/N$ asymptotic behavior~\eqref{eq:sf.P} and~\eqref{eq:sf}.
It is straightforward to determine $a_N$ and $b_N$. Let $\overline{n}$ be a positive parameter of order one: $\overline{n} \sim O(N^0)$.
We have 
\begin{equation}
    P_+(x)=e^{-x}=\frac{\overline{n}}{N}\qquad\Rightarrow\qquad x=\ln N-\ln \overline{n}.
\end{equation}
We identify the scaling in this non-standard way compared to most of the literature, where one usually approaches it via the limit of the distribution of the leading particle because our choice has a very natural interpretation; namely, $\overline{n}$ can be understood as the average fraction of particles in the leading positions. It does not matter which kind of limiting extreme value statistics one encounters, whether given by Gumbel, Fréchet, or Weibull distributions, 
so this approach serves as a good starting point when generalizing the results to other types of statistics.

In this article, we only consider distributions belonging to the Gumbel class from the point of view of extreme values, for which the relation $\zeta=-\ln \overline{n}$, or equivalently $\overline{n}=e^{-\zeta}$, always holds. The cumulative probability distribution of the leader is
\begin{equation}
    {\rm Prob}(x_{max,N}<x)  \overset{N\gg1}{\approx} \exp(- N P_+(x))\approx \exp(- \overline{n})=\exp(- e^{-\zeta}) ,
\end{equation}
{\em cf.}~\eqref{eq:ev}.
The new variable $\zeta$ is also of order one. Comparing with~\eqref{eq:scaling.position}, we can identify the scaling parameters
\begin{equation}\label{scaling.constant.drift}
    a_N= \ln N\qquad{\rm and}\qquad b_N=1.
\end{equation}
Substituting these values into~\eqref{eq:qpp_drift}, we notice that the time does not need to be rescaled to have~\eqref{eq:sf}. We use a numerical value $c_N=8$ to simplify the limit expression for the overlap ratio; 
see Eq. \eqref{eq:omega_inf} given in the introduction. 
Summarizing, the original parameters $y,x,t$ are rescaled to $\xi,\zeta,\tau$ 
as follows
\begin{equation}
y = \ln N + \xi; \ x=\ln N + \zeta; \ t = 8\tau .
\label{eq:scaling_drift}
\end{equation}

In the leading order, we find
\begin{equation}
    P_+(y) = \frac{1}{N} \widehat{P}_+(\xi) = \frac{1}{N} e^{-\xi},\qquad P_+(x) = \frac{1}{N} \widehat{P}_+(\zeta) = \frac{1}{N} e^{-\zeta}    \qquad{\rm and}\qquad W(y,x,t)=\widehat{W}(\xi,\zeta,\tau) \approx \frac{e^{-\frac{( (\zeta-\xi)+4\tau)^2}{16 \tau}}}{\sqrt{16\pi \tau}} 
    \label{eq:W16}
\end{equation}
The last approximation becomes clear when noticing that the first term on the right-hand side of~\eqref{eq:Wd} is of order one 
because $\ln N$ cancels in the difference $x-y = \xi-\zeta$, while the other two terms vanish for $N\rightarrow \infty$. 
The two-point probability density can be calculated from the propagator
\eqref{eq:q} yielding
\begin{equation}
  \widehat{q}(\xi,\zeta,\tau) \approx \frac{e^{-\frac{(\zeta-\xi)^2}{16 \tau} - \frac{1}{2}(\xi+\zeta) - \tau}}{\sqrt{16\pi \tau}} .
  \label{eq:qq}
\end{equation}
The cumulative distribution function can be calculated either from~\eqref{eq:qpp_drift} or by integrating~\eqref{eq:qq},
\begin{equation}
    \widehat{Q}_{++}(\xi,\zeta,\tau) \approx
    \frac{1}{2} e^{-\frac{\xi+\zeta}{2}}
    \left(e^\frac{\xi-\zeta}{2} {\rm erfc}\left(\frac{\xi-\zeta+4\tau}{4\sqrt{\tau}}\right)+
    e^\frac{\zeta-\xi}{2} {\rm erfc}\left(\frac{\zeta-\xi+4\tau}{4\sqrt{\tau}}\right)  \right) . 
    \label{eq:NQpp}
\end{equation}
Looking at the above equations, we can see that one can isolate contributions from the difference $\beta=\xi-\zeta$
and the center of mass $\alpha = (\xi+\zeta)/2$. 

We can now use the integration measure \eqref{eq:qq} and the cumulative distribution function \eqref{eq:NQpp}
to find the generating function $\widehat{P}_R(z,w,\tau)$ \eqref{eq:PRmain} to leading order for large $N$.
Note that the integration measure \eqref{eq:qq} is not normalizable in this approximation because it diverges for 
$\alpha =(\xi+\zeta)/2 \rightarrow -\infty$. 
This divergence does not cause any problems because a different exponential term introduces a strong damping factor that makes the integral well-defined.
From a physical point of view, it is clear why $\alpha\to-\infty$ does not play any role, since it corresponds to the left tail of the particle distribution, which has no effect on the leading particles located in the right tail of the distribution.
The whole expression after inserting~\eqref{eq:qq} and~\eqref{eq:NQpp} into \eqref{eq:PRmain} and performing the change of variables to $\alpha$ and $\beta$ is
\begin{equation}
\begin{split}
  \widehat{P}_R(z,w,\tau) \approx 
  \frac{1}{\sqrt{16 \pi \tau }}
  \int_{-\infty}^{+\infty} d\alpha e^{-\alpha} \int_{-\infty}^{+\infty} d\beta 
   e^{-\frac{1}{16\tau}\beta^2}   
   \exp\left(-(1-z) e^{-\alpha-\frac{\beta}{2}} - (1-w) e^{-\alpha+\frac{\beta}{2}}  \right. \\ \left. +
   \frac{(1-z)(1-w)}{2} e^{-\alpha}
    \left(e^\frac{\beta}{2} {\rm erfc}\left(\frac{\beta+4\tau}{4\sqrt{\tau}}\right)+ 
    e^{-\frac{\beta}{2}} {\rm erfc}\left(\frac{-\beta+4\tau}{4\sqrt{\tau}}
   \right)\right)\right) .
\end{split}
\end{equation}
The integral over $\alpha$ can actually be calculated. 
After changing $s=\beta/(4\sqrt{\tau})$, we obtain 
\begin{equation}
 \widehat{P}_R(z,w,\tau) \approx 
  \frac{e^{-\tau}}{\sqrt{\pi}(1-z)(1-w)}
   \int_{-\infty}^{+\infty} ds 
   \frac{ e^{-s^2}}{ e^{+2s\sqrt{\tau}}\left(\frac{1}{1-z} - \frac{1}{2}  {\rm erfc}\left[\sqrt{\tau}+s
   \right]\right) + e^{-2s\sqrt{\tau}}\left(\frac{1}{1-w} - \frac{1}{2}  {\rm erfc}\left[\sqrt{\tau}-s
   \right]\right)
   } .
   \label{eq:PRS}
\end{equation}
This is an explicit expression for the generating function. It can be used to calculate 
the reshuffling probabilities \eqref{eq:pRres} or the overlaps \eqref{eq:Z}. 

Setting $z=w=0$ in \eqref{eq:PRS}, we find the probability that the leader remains the leader after a time $t=8\tau$,
\begin{equation}
  \langle \widehat{\Omega}_1(\tau)\rangle = \widehat{p}_R(1,1,\tau) \approx 
   \frac{e^{-\tau}}{\sqrt{\pi}} 
   \int_{-\infty}^{+\infty} ds 
   \frac{ e^{-s^2}}{ e^{+2s\sqrt{\tau}}\left(1 - \frac{1}{2} {\rm erfc}\left[\sqrt{\tau}+s
   \right]\right) + e^{-2s\sqrt{\tau}}\left(1 - \frac{1}{2} {\rm erfc}\left[\sqrt{\tau}-s
   \right]\right)
   } .
\end{equation}
For $\tau=0$, the integral is equal to $1$, as it should be. 
  For any $n>1$, the average overlap  ratio~\eqref{eq:OmegaZ} becomes~\cite{BK}
\begin{equation}
\langle \widehat{\Omega}_n(\tau)\rangle \approx \frac{1}{n} \int_{-\infty}^{+\infty} \frac{ds e^{-s^2}}{\sqrt{\pi}} 
\sum_{j=1}^n\sum_{k=1}^n (-)^{j+k} \binom{j+k-2}{j-1}
 \binom{n}{j} \binom{n}{k}  e^{(j-k) 2s\sqrt{t}} \left(\frac{1}{f(s,\tau)}\right)^{j+k-1} ,
 \label{eq:Omega}
\end{equation}
where
\begin{equation}
    f(s,\tau) = e^{+2s\sqrt{\tau}}\left(1 - \frac{1}{2} {\rm erfc}\left[\sqrt{\tau}+s
   \right]\right) + e^{-2s\sqrt{\tau}}\left(1 - \frac{1}{2} {\rm erfc}\left[\sqrt{\tau}-s
   \right]\right) .
\end{equation}
It is essentially a one-dimensional Gaussian integral of a function whose complexity increases with $n$, as the following formulae for $n=1,2,3$ show
\begin{equation}
\begin{split}
    \langle \widehat{\Omega}_1(\tau)\rangle & \approx \int_{-\infty}^{+\infty} \frac{ds e^{-s^2}}{\sqrt{\pi}}  \frac{1}{f(s,\tau)}, \\
    \langle \widehat{\Omega}_2(\tau)\rangle & \approx  \int_{-\infty}^{+\infty} \frac{ds e^{-s^2}}{\sqrt{\pi}} \left( \frac{1}{f^3(s,\tau)} -  \frac{e^{2s\sqrt{\tau}}+e^{-2s\sqrt{\tau}}}{f^2(s,\tau)} 
       +\frac{2}{f(s,\tau)}\right), \\
    \langle \widehat{\Omega}_3(\tau)\rangle & \approx \int_{-\infty}^{+\infty} \frac{ds e^{-s^2}}{\sqrt{\pi}} \left( \frac{2}{f^5(s,\tau)}- \frac{3 e^{2s\sqrt{\tau}}+ 3 e^{-2s\sqrt{\tau}}}{f^4(s,\tau)} +
    \frac{6 + e^{4s\sqrt{\tau}}+e^{-4s\sqrt{\tau}}}{f^3(s,\tau)} - \frac{3 e^{2s\sqrt{\tau}}+ 3 e^{-2s\sqrt{\tau}}}{f^2(s,\tau)} + \frac{3}{f(s,\tau)}\right) .
\end{split}
\end{equation}
The integrals can be computed numerically up to $n$ of order ten. Remarkably, in the limit $n\rightarrow \infty$, the expression \eqref{eq:Omega}
drastically simplifies to~\eqref{eq:omega_inf}, 
as shown in~\cite{BK}.

\section{Ornstein-Uhlenbeck process \label{sec:OU}}

With the knowledge from the previous section, we move on to diffusion in a quadratic potential $V(x)=\frac{k}{2} x^2$
with a diffusion coefficient $D=\frac{\sigma^2}2$, which was numerically investigated in~\cite{BK}. Like before, we set $k=1$ and $\sigma=1$ without loss of generality so that the Fokker-Planck equation~\eqref{eq:FP} becomes 
\begin{equation}
    \partial_t p(x,t) =  \partial_x (x p(x,t)) + \frac{1}{2} \partial^2_{xx} p(x,t).
    \label{eq:OU1}
\end{equation}
As discussed at the end of Section~\ref{sec:prelim}, using the transformation \eqref{eq:transf},
it is easy to recover a general form of this equation corresponding to any $k$ and $\sigma$. The Green function
(heat kernel or propagator) for the Ornstein-Uhlenbeck equation~\eqref{eq:OU1} is \cite{R} 
\begin{equation}
  W(y,x,t) = \frac{1}{{\sqrt{\pi (1-e^{-2t})}}} 
  \exp\left( - \frac{(x-ye^{-t})^2}{(1-e^{-2t})} \right) = \frac{1}{\sqrt{\pi (1-\rho^2)}} 
  \exp\left( -\frac{(x-\rho y)^2}{(1-\rho^2)} \right)
\end{equation}
when we employ the auxiliary variable 
\begin{equation}
    \rho = e^{-t} .
    \label{eq:rho}
\end{equation}
Due to the exponential suppression of the initial position for large times, the probability density function of the stationary distribution is 
\begin{equation}
  p(x) = W(y,x,t \rightarrow \infty) = \frac{1}{\sqrt{ \pi}} e^{-x^2}.
  \label{eq:pg}
\end{equation}
Combining this with the propagator, we obtain the two-point distribution function~\eqref{eq:q},
\begin{equation}
    q(y,x,t) = \frac{1}{\pi \sqrt{1-\rho^2}} 
    \exp\left( -\frac{y^2 +x^2-2\rho xy}{(1-\rho^2)} \right) .
    \label{eq:qOU}
\end{equation}
The parameter $\rho$ plays the role of the Pearson correlation coefficient between $y$ and $x$. 
For large times, $\rho\rightarrow \infty$ and the positions completely de-correlate into two independent Gaussian variables.
The complementary cumulative distribution function of this stationary probability 
distribution~\eqref{eq:pg} is the complementary error function
\begin{equation}
   P_+(x) = \frac{1}{\sqrt{\pi}}\int^{+\infty}_{x} d u e^{-u^2} = \frac{1}{2} {\rm erfc}\left(x\right) \overset{x\gg1}{\approx} \frac{1}{\sqrt{4\pi} x} e^{-x^2} .
\end{equation}
As before, we determine the scale of the variable $x$, especially the scaling coefficients $a_N$ and $b_N$ in~\eqref{eq:scaling.position}, by setting 
\begin{equation}\label{OU-scale-determination}
   P_+(x) \overset{x\gg1}{\approx} \frac{1}{\sqrt{4\pi} x} e^{-x^2}=\frac{\overline{n}}{N} 
\end{equation}
with $\overline{n}$ being a positive variable of order one. We can rearrange this equation to an implicit equation for $x$
\begin{equation}
   x=\sqrt{\ln N}\sqrt{1-\frac{\ln x}{\ln N}-\frac{\ln \overline{n}}{\ln N}-\frac{\ln (4\pi)}{2\ln N}}.
\end{equation}
As $1\ll\ln(\ln(N))\ll\ln N$, we see that the leading order of $x$ is $\sqrt{\ln N}$. Reinserting $x$ and expanding up to order $\ln \overline{n}/\sqrt{\ln N}$, as all other powers are much smaller, we find
\begin{equation}
   x=\sqrt{\ln N}-\frac{\ln (\ln N)}{4\sqrt{\ln N}}-\frac{\ln (4\pi)}{4\sqrt{\ln N}}-\frac{\ln \overline{n}}{2\sqrt{\ln N}}+o\left(\frac{1}{\sqrt{\ln N}}\right).
\end{equation}
By identifying $\xi = - \ln \overline{n}$ and comparing it with \eqref{eq:scaling.position}, we can find the scaling parameters \eqref{eq:scaling.position}
\begin{equation}
    a_N=\sqrt{\ln N}-\frac{\ln (\ln N)}{4\sqrt{\ln N}}-\frac{\ln (4\pi)}{4\sqrt{\ln N}}\qquad{\rm and}\qquad b_N=-\frac{1}{2\sqrt{\ln N}}\approx\frac{1}{2a_N}.
    \label{eq:aN}
\end{equation}
This scaling could have been obtained when fine-tuning the scaling for the cumulative distribution of the leading particle to be the Gumbel distribution~\cite{Gumbel} for  $N\rightarrow \infty$, {\em i.e.},
\begin{equation}
    \lim_{N\to\infty}{\rm Prob}(y_{max,N}<a_N+b_N \xi) = e^{-e^{-\xi}}
\end{equation}
at time $t_1$, and $e^{-e^{-\zeta}}$ at time $t_2$.

In conclusion, the leaders at 
times $t_1$ and $t_2$ are located at 
\begin{equation}
    y = a_N + \frac{\xi}{2 a_N}; \quad x = a_N + \frac{\zeta}{2 a_N}
    \label{eq:scale}
\end{equation}
where $\xi$ and $\zeta$ are of order unity and $a_N$ is given in~\eqref{eq:aN}.
From that, we can find the rescaled function~\eqref{eq:sf.P} which is in the present case
\begin{equation}
\begin{split}
  P_+(y)  = \frac{1}{N} \widehat{P}_+(\xi) +o(N^{-1}) = \frac{1}{N} e^{-\xi}+o(N^{-1})\qquad{\rm and}\qquad P_+(x)  = \frac{1}{N} \widehat{P}_+(\zeta) +o(N^{-1}) = \frac{1}{N} e^{-\zeta}+o(N^{-1}).
\end{split}  
\end{equation}
When substituting the variables~\eqref{eq:scale} in the density of the two-point function~\eqref{eq:q}, we obtain for the rescaled function~\eqref{eq:sf}
\begin{equation}
    \widehat{q}(\xi,\zeta,t) \approx \frac{N}{4 a_N^2 \pi\sqrt{1-\rho^2} } 
    \exp\left( -\frac{1}{1-\rho^2}\left( 2 (1-\rho) a^2_N  + (1-\rho)(\xi+\zeta) +
    \frac{1}{4 a_N^2} (\xi^2 - 2\rho \xi\zeta + \zeta^2) \right)\right).
    \label{eq:Nq}
\end{equation}
The factor $(4a_N^2)^{-1}$ in the numerator of the first fraction comes from the Jacobian $dydz = (4a_N^2)^{-1}d\xi d\zeta$.
Substituting $N$ in the last equation with $\sqrt{4\pi} a_N e^{a_N^2}$, see~\eqref{OU-scale-determination} for $x=a_N$ corresponding to $\overline{n}=1$, we obtain
\begin{equation}
    \widehat{q}(\xi,\zeta,t) \approx \frac{1}{\sqrt{4\pi} a_N\sqrt{1-\rho^2}} 
    \exp\left( -\frac{1-\rho}{1+\rho} a_N^2 - \frac{\xi+\zeta}{1+\rho} -
    \frac{1}{4(1-\rho^2) a_N^2} (\xi^2 - 2\rho \xi\zeta + \zeta^2) \right).
    \label{eq:Nq1}
\end{equation}
For any constant value of $\rho=e^{-t} \in [0,1)$, the last term in the exponent 
goes to zero for $N\rightarrow \infty$ because $1/a_N^2 \rightarrow 0$.
Thus, let us at the moment neglect this term and use the relation between $a_N$ and $N$ in~\eqref{eq:aN}. Then, the following asymptotic behavior holds true for large $N$
\begin{equation}
    \widehat{q}(\xi,\zeta,t) \approx \frac{1}{\sqrt{4\pi}}\frac{1}{\sqrt{1-\rho^2}}  
    e^{-\frac{1}{1+\rho}(\xi+\zeta)} \Phi(N) 
    \label{eq:Nqa}
\end{equation}
and 
\begin{equation}
\widehat{Q}_{++}(\xi,\zeta,t) = \int_\xi^\infty d\xi' \int_\zeta^\infty d\zeta' \widehat{q}(\xi',\zeta',t)  = 
 \frac{1}{\sqrt{4\pi}} \frac{(1+\rho)^2}{\sqrt{1-\rho^2}} e^{-\frac{\xi+\zeta}{1+\rho}} \Phi\left(N\right)
\label{eq:Qhat}
\end{equation}
where
\begin{equation}
\Phi(N) =a_N^{-1}\exp\left[-\frac{1-\rho}{1+\rho} a_N^2\right]\approx\left(\frac{N}{\sqrt{4\pi}}\right)^{-\frac{1-\rho}{1+\rho}} 
         \left(\ln \frac{N}{\sqrt{4\pi}}\right)^{-\frac{\rho}{1+\rho}} .
         \label{eq:Phi}
\end{equation}
This result was first obtained in \cite{MS}.

We see that for any finite value of $\rho=e^{-t} \in [0,1)$, the scaling factor $\Phi(N)\rightarrow 0$ vanishes
when $N\rightarrow \infty$, which means that for any finite time $t$, the probability that a leader 
will still be among the leading particles after time $t$ vanishes for $N\rightarrow \infty$. 
This can be interpreted as a rapid mixing of particles. In other words, leaders quickly cease to be leaders, 
and particles with a low ranking can quickly become leaders when the population size $N$ increases.

The situation changes in the double scaling limit: $\rho = e^{-t} \rightarrow 1$ for $N\rightarrow \infty$. 
In particular, the two-point correlation function~\eqref{eq:Nq1} will be of order one when $N\rightarrow \infty$ if we set
\begin{equation}
   \rho=e^{-t}= 1 - \frac{2\tau}{a_N^2} \approx 1 - \frac{2\tau}{\ln N} 
   \label{eq:rhoN}
\end{equation}
where $\tau$ is an $N$-independent rescaled time. This corresponds to the timescale 
\begin{equation}
  t =c_N \tau =\frac{2\tau}{a_N^2}\approx \frac{2\tau}{\ln N} \qquad\Rightarrow\qquad c_N =\frac{2}{a_N^2}\approx \frac{2}{\ln N}.
  \label{eq:ttau}
\end{equation} 
The factor of two in the numerator is chosen for convenience. 
When $\rho$ is given by \eqref{eq:rhoN}, the last term in the exponent \eqref{eq:Nq1} can be approximated for
$N\rightarrow \infty$ by 
\begin{equation}
    \frac{1}{4(1-\rho^2) a_N^2}\left(\xi^2 - 2\rho\xi\zeta + \zeta^2\right) \approx \frac{1}{16\tau} (\xi-\zeta)^2 
\end{equation}
as follows from the identity 
\begin{equation}
    \xi^2 - 2\rho \xi\zeta + \zeta^2 = \frac{1+\rho}{2} (\xi-\zeta)^2 
    + \frac{1-\rho}{2}(\xi+\zeta)^2 .
\end{equation}
The remaining terms in \eqref{eq:Nq1} also simplify in the limit $N\rightarrow \infty$, yielding 
\begin{equation}
    \widehat{q}(\xi,\zeta,\tau) \approx \frac{1}{\sqrt{16\pi \tau }}
e^{-\frac{1}{16\tau}(\xi-\zeta)^2-\frac{\xi+\zeta}{2} -\tau} .
\label{eq:Nqas}
\end{equation}
This measure is identical to the asymptotic measure \eqref{eq:qq} that we obtained for the constant drift diffusion. Obviously, 
$\widehat{Q}_{++}(\xi,\zeta,\tau) = \int_{\xi}^\infty \int_{\zeta}^\infty du dv \widehat{q}(u,v,\tau)$ is also identical to \eqref{eq:NQpp},
and as a consequence, we obtain exactly the same formula for the reshuffling probability generating function as for the diffusion with a constant drift
and a reflecting wall \eqref{eq:PRS}. 

Consequently, for large $N$, all reshuffling probabilities are identical to those for diffusion with a constant drift in the presence
of a reflecting wall. Two of the authors have numerically observed this in~\cite{BK}, but due to the large finite size correction, we could not definitively say that it was true. The only difference from Brownian motion with a drift to a reflective wall  is the scaling as a function of population size $N$.
In the case of diffusion with a constant drift (in a linear potential), the scaling equations are given by \eqref{eq:scaling_drift}, 
while in the case of diffusion in a quadratic potential – by \eqref{eq:scale} and \eqref{eq:ttau}.
The timescale of leaders reshuffling in a quadratic potential decreases logarithmically with 
the population size $t \sim 1/\ln N$, while in a linear potential it is asymptotically independent of the population size. 

\section{Universality\label{sec:univ}}

Next, we move on to diffusion in potentials that 
behave asymptotically as $V(x) \sim \frac{1}{2} x^\gamma$ with $\gamma>0$ for $x\to+\infty$. We assume the existence of a stationary state and focus on
the behavior of the system in the stationary state.
In this case, the drift behaves asymptotically as 
$V'(x) \sim \frac{\gamma}{2} x^{\gamma-1}$, so that the corresponding diffusion equation (for $\sigma=1$) asymptotically takes the form
\begin{equation}
  \partial_t p(x,t) = \frac{\gamma}{2} \partial_x (x^{\gamma-1} p(x,t)) + 
  \frac{1}{2} \partial_{xx}^2 p(x,t) ,
    \label{eq:FPgamma}
\end{equation}
for $x\rightarrow \infty$.
Thence, the probability density of the stationary state is asymptotically
\begin{equation}
    p(x)=ce^{-2V(x)} \overset{x\gg1}{\approx} c e^{-x^\gamma}
\end{equation}
with the normalization constant $c$. The complementary cumulative distribution function becomes
\begin{equation}
    P_+(x)=\int_x^\infty ce^{-2V(u)}d u=\int_0^\infty c\exp\left[-2V\left(x+\frac{u}{2V'(x)}\right)\right]\frac{du}{2V'(x)} \overset{x\gg1}{\approx}\frac{c}{2V'(x)}e^{-2V(x)}\approx \frac{c}{\gamma x^{\gamma-1}} e^{-x^\gamma} ,
\end{equation}
where we substituted $u \to x+u/(2V'(x))$ and then Taylor expanded it to the first order. This computation highlights what is a general condition for the validity of the expansion, which requires $\lim_{x\to+\infty}V'(x)x=+\infty$ to ensure $x\gg1/V'(x)$. This condition is fulfilled for the considered potentials.

Following the method described in Sections~\ref{sec:const_drift} and~\ref{sec:OU}, we find the proper scaling of the position of the leading particles by setting
\begin{equation}
    P_+(x)\overset{x\gg1}{\approx}\frac{c}{\gamma x^{\gamma-1}} e^{-x^\gamma}=\frac{\overline{n}}{N}
\end{equation}
with a positive variable $\overline{n}>0$ that is of order one for 
large $N$. The last equation can be rewritten as  
\begin{equation}
    x=(\ln N)^{1/\gamma}\left[1-(\gamma-1)\frac{\ln x}{\ln N}-\frac{\ln \overline{n}}{\ln N}-\frac{\ln\left(\gamma /c\right)}{\ln N}\right]^{1/\gamma},
\end{equation}
which helps us identify the leading contribution $(\ln N)^{1/\gamma}$ of the scaling and provides an expansion because $1\ll\ln(\ln N)\ll\ln N$. We arrive at
\begin{equation}
    x=(\ln N)^{1/\gamma}+\frac{1-\gamma}{\gamma^2}\frac{\ln(\ln N)}{(\ln N)^{(\gamma-1)/\gamma}}-\frac{1}{\gamma}\frac{\ln(\gamma/c)}{(\ln N)^{(\gamma-1)/\gamma}}-\frac{1}{\gamma}\frac{\ln\overline{n}}{(\ln N)^{(\gamma-1)/\gamma}}+o\left(\frac{1}{(\ln N)^{(\gamma-1)/\gamma}}\right)
\end{equation}
leading to the identification $\zeta=-\ln\overline{n}$ and
\begin{equation}
\begin{split}
    a_N =(\ln N)^{1/\gamma}+\frac{1-\gamma}{\gamma^2}\frac{\ln(\ln N)}{(\ln N)^{(\gamma-1)/\gamma}}-\frac{1}{\gamma}\frac{\ln(\gamma/c)}{(\ln N)^{(\gamma-1)/\gamma}} \overset{N\gg1}{\approx} (\ln N)^{1/\gamma} \qquad{\rm and}\qquad
    b_N =\frac{(\ln N)^{(1-\gamma)/\gamma}}{\gamma } \approx \frac{1}{\gamma a_N^{\gamma-1}},
\end{split}
\label{eq:abN}
\end{equation}
see \eqref{eq:scaling.position}. This scaling naturally reduces to~\eqref{scaling.constant.drift} for $\gamma=1$ (diffusion with constant drift, where $c=1$) and to~\eqref{eq:aN} for $\gamma=2$ (Ornstein-Uhlenbeck process, which is the diffusion in a harmonic trap, where $c=1/\sqrt{\pi}$).

The next step is to determine the dependence of the coefficient $c_N$\eqref{eq:scaling.time} on the population size $N$, in order 
to establish the correct timescale for the leaders' reshuffling.
As discussed in section \ref{sec:scaling}, a way to achieve this goal is to look for a timescale on which the cumulative distribution 
function $Q_{++}$ and the integration measure $q$ behave asymptotically as $1/N$ for large $N$; see equation \eqref{eq:sf}.
The measure can be determined from the propagator: $q(y,x,t) = p(y)W(y,x,t)$ \eqref{eq:q}, and the propagator can be calculated using Eq. \eqref{eq:W}.
The equation for the propagator can be solved asymptotically for $y,x\rightarrow \infty$, which is sufficient for our purposes because we are interested in
the leaders. For large $x,y$, Eq. \eqref{eq:W} takes the following asymptotic form
\begin{equation}
  \partial_t W(y,x,t) = \frac{\gamma}{2} \partial_x (x^{\gamma-1} W(y,x,t)) +
  \frac{1}{2} \partial_{xx}^2 W(y,x,t).
    \label{eq:Wgamma}
\end{equation}
If we express the propagator $W(x,y,t)$ in terms of the scaled parameters 
$\widehat{W}(\xi,\zeta,\tau)$, where $x=a_N \xi + b_N$, $y=a_N \zeta + b_N$, 
and $t= c_N \tau$, the last equation can be written as
\begin{equation}
  \frac{1}{c_N} \partial_\tau \widehat{W}(\xi,\zeta,\tau) =  \frac{\gamma^2 a_N^{2\gamma-2}}{2} \left[\partial_\zeta \widehat{W}(\xi,\zeta,\tau) +   
   \partial_{\zeta\zeta}^2 \widehat{W}(\xi,\zeta,\tau)\right]
    \label{eq:WHgamma}
\end{equation}
in the leading order, as follows from \eqref{eq:abN}. We have neglected the term 
\begin{equation}
    \frac{\gamma(\gamma-1)a_N^{\gamma-2}}{2} \partial_\zeta\left[\zeta \widehat{W}(\xi,\zeta,\tau)\right]
    \label{eq:curv}
\end{equation}
because the coefficient in front of this term increases with $N$ like 
$(\ln N)^\frac{\gamma-2}{\gamma}$, that is, more slowly than the coefficients of the other terms~\eqref{eq:WHgamma}. The derivative $\partial_\zeta$ in \eqref{eq:curv}
cannot change the dependence on $N$, as the term is neither strongly oscillating nor rapidly decreasing/increasing in $N$.

Comparison of~\eqref{eq:WHgamma} with the Fokker-Planck equation~\eqref{FP.eq.constant.drift} for diffusion with constant drift reveals that it becomes the same equation (taking into account that there $t=8\tau$) when setting the time scaling factor equal to
\begin{equation}\label{time.scale.general}
    c_N=\frac{8}{\gamma^2 a_N^{2\gamma-2}}\overset{N\gg1}{\approx}\frac{8}{\gamma^2}(\ln N)^{(2-2\gamma)/\gamma} .
\end{equation}
Then, the asymptotic equation~\eqref{eq:WHgamma} describes an asymptotically free diffusion with a constant drift equal to $-1/2$,
$\sigma=1$, and time $t=8\tau$, exactly as equation~\eqref{FP.eq.constant.drift}. Solving this equation, we find 
the propagator~\eqref{eq:W16}, and after using~\eqref{eq:q}, we recover the measure~\eqref{eq:qq} and the cumulative distribution function \eqref{eq:NQpp}, which are identical to those for diffusion with constant drift.

Therefore, we obtain the same generating function~\eqref{eq:PRS}, the same
overlap $\langle \widehat{\Omega}_\infty(\tau)\rangle = {\rm erfc}(\sqrt{\tau})$, 
{\em etc.}, independently of the power $\gamma$.
The solution is universal for diffusion in this class of potentials, which means that all the reshuffling probabilities between
leaders are independent of $\gamma$ in the limit of an infinitely large population,
$N\rightarrow \infty$, once a proper rescaling has been performed.

Finally, it is worth noting that the range of fluctuations in the leaders’ positions, expressed in units of their typical positions, approaches zero asymptotically as $b_N/a_N \sim (\ln N)^{-1}$ when $N$ increases, regardless of $\gamma$.

We also note that $c_N$ is asymptotically proportional to $b_N^2$ for large $N$, again independently of $\gamma$. 
This can be understood as follows. The typical distance between the positions 
of leaders having two consecutive ranks scales linearly with $b_N$ \cite{SM}, so the time needed for a random walker to cover this distance 
scales quadratically with $b_N$.

\section{Going beyond: free diffusion \label{sec:free}}

In this section, we address ranking statistics for free diffusion with a 
normal (Gaussian) initial condition. 
This case was also studied in~\cite{LeD}. The main difference from the cases we 
have discussed so far is that there is no stationary state. Yet, there is something similar, namely shape preservation, though the scales are changing. If the
initial distribution of particle positions is given by a Gaussian distribution, 
the distribution will remain Gaussian for all times. The only parameter that changes is the width of the Gaussian distribution, 
which grows with time. The positions of the leading particles 
will be time-dependent as well.

The Fokker-Planck equation is simply the one-dimensional heat equation 
without a drift term
\begin{equation}
    \partial_t p(x,t) = \frac{1}{2}\partial_{xx}^2 p(x,t).
\end{equation}
Choosing a fully localized state $p(x,0) = \delta(x)$ at $t=0$, the probability density function will be Gaussian at all times $t>0$,
\begin{equation}\label{free.pdf}
p(x,t) = \frac{1}{\sqrt{2\pi t}} e^{-\frac{x^2}{2 t}}.
\end{equation}
We could have also started from a time $t_1$ with a centered Gaussian having a standard deviation $\sqrt{t_1}$ and understood \eqref{free.pdf} as an evolved distribution after time $t=t_2-t_1$.

The heat kernel for $t_2>t_1$ is
\begin{equation}
W(y,t_1; x,t_2) = \frac{1}{\sqrt{2\pi (t_2-t_1)}} 
e^{-\frac{(y-x)^2}{2 (t_2-t_1)}} .
\end{equation}
Because free diffusion is nonstationary, we need to keep the dependence on $t_1$ and $t_2$ in the expressions and not on the time difference.
The probability density function of the two-point probability distribution is
\begin{equation}
  q(y,t_1;x,t_2) = p(y,t_1) W(y,t_1;x,t_2) = 
  \frac{1}{2\pi  \sqrt{t_1  (t_2-t_1)}} 
  \exp\left( -\frac{1}{2} \frac{y^2 t_1 + x^2 t_2 - 2 t_1 xy }{t_1(t_2-t_1)}\right) .
\end{equation}
Now we can use the fact that rescaling all particle positions by a common factor does not change their ranking. 
So let us apply the following rescaling:
\begin{equation}
    \widetilde{x}= \frac{x}{\sqrt{2 t_1}}\qquad {\rm and}\qquad \widetilde{y}= \frac{y}{\sqrt{2 t_2}} .
\end{equation}
Both density functions at $t_1$ and $t_2$ of the rescaled variables $\widetilde x$ and $\widetilde y$ are identical
\begin{equation}
    \widetilde{p}(\widetilde{x},t_1) = \widetilde{p}(\widetilde{x})= \frac{1}{\sqrt{\pi}} e^{-\widetilde{x}^2}\qquad {\rm and}\qquad \widetilde{p}(\widetilde{y},t_2) = \widetilde{p}(\widetilde{y})= \frac{1}{\sqrt{\pi}} e^{-\widetilde{y}^2}.
\end{equation}
Hence, they look like a stationary state under this time-dependent rescaling.

The two-point function in these rescaled variables is equal to
\begin{equation}
  \widetilde{q}(\widetilde{y},t_1;\widetilde{x},t_2) = 
  \frac{1}{\pi\sqrt{1-\widetilde{\rho}^2}}
  \exp\left( -\frac{\widetilde{y}^2 + \widetilde{x}^2 - 2 \widetilde{\rho} \widetilde{x} \widetilde{y}}{1-\widetilde{\rho}^2}\right)
  \label{eq:qt}
\end{equation}
where 
\begin{equation}
    \widetilde{\rho} = \sqrt{\frac{t_1}{t_2}} .
    \label{eq:rhot}
\end{equation}
Comparing~\eqref{eq:qt} and~\eqref{eq:qOU}, we see that the expressions in the rescaled variables are identical to those for 
the stationary Ornstein-Uhlenbeck process, except that the correlation coefficient $\rho$ \eqref{eq:rho} is replaced with $\widetilde{\rho}$ \eqref{eq:rhot}. 
Apart from this minor difference, the rank reshuffling dynamics in both cases are therefore identical. The relationship between physical time and the
rescaled time $\tau$, where one observes the scaling relations described in the previous section, see~\eqref{eq:rhoN}, is 
\begin{equation}
    \sqrt{\frac{t_1}{t_2}} =\sqrt{\frac{t_1}{t_1+t}} = 1 - \frac{2\tau}{a^2_N} \approx 1 - \frac{2\tau}{\ln N}\qquad \Leftrightarrow\qquad t\approx\frac{4t_1}{\ln N}\tau.
\end{equation}
Hence, the scaling constant $c_N=4t_1/\ln N$ for the time $t$ is now also dependent on the initial condition, here in the form of the initial time $t_1$, which is related to the width of the initial Gaussian distribution. Surprisingly, the reshuffling rate for the leading particles is faster than inside a potential $V(x)\sim x^\gamma/2$ for $x\to+\infty$ with $\gamma<2$. This was already observed in~\cite{LeD}. Actually, the time scaling for the confining potential blows up when $\gamma\to0$; see~\eqref{time.scale.general}. Thus, something dramatic is happening when going from stationary states in confining potentials to non-stationary ones when the potential is not confining.

\section{Conclusions \label{sec:concl}}

In this paper, we derived an explicit expression for the generating function for the probabilities describing the reshuffling of the order 
statistics of particles performing Brownian motion in the stationary state in a confining potential that asymptotically behaves as $V(x) \propto x^\gamma$, where $\gamma> 0$ 
for $x\rightarrow+\infty$. We have demonstrated the universality of the result by showing
that the generating function, and hence the ranking dynamics, are independent of $\gamma$ 
if we properly rescale time $t \sim \left(\ln N\right)^{2(\gamma-1)/\gamma} \tau$ and express it in terms of an effective time variable $\tau$ that is of order one. The universality can be understood 
when one considers the positions of the highest-ranking particles on a local scale, which is magnified locally
so that, on the one hand, the potential $V(x)$ can be linearly approximated on this scale, and on the other hand, there are suitably many high-ranking particles
in the magnified range. This certainly requires that the width of the individual distribution of the leading particles is much smaller than the position where they can be found. Going beyond this, some deviations from our results might occur and remain open.

But also for finite $N$ deviations from linearity arise, manifesting as curvature of the potential, which leads 
to larger finite size corrections to the limiting expressions compared to Brownian motion with a constant drift towards a reflective wall, as found in Ref.~\cite{BK}.
This curvature is encoded in terms like those in~\eqref{eq:curv}, for example. They can be neglected only in the limit $N\rightarrow \infty$.
The challenge is to establish the dependence of the finite-size corrections on $\gamma$ and $N$ so that they can be controlled systematically. In particular, in numerical simulations, these corrections may overshadow the universal results derived here. Therefore, it might be of practical interest to quantify them.

The universality class discussed in this paper also includes 
free diffusion with a normal initial condition, as it can be mapped 
by a simple rescaling of the width of the normal distribution 
to a stationary picture that is isomorphic to the stationary 
state of diffusion in a quadratic potential. The initial condition is not extremely restrictive, as for long enough waiting times, any distribution would smooth out and appear Gaussian. What is, however, puzzling in this setting is that the scaling of the time difference is of order $1/\ln N$, exactly the same as that of the Ornstein-Uhlenbeck process. This is surprising, as the time scale in the studied confining potentials diverges to infinity when taking the limit of the parameter $\gamma\to0$, which is another way to approach the free diffusion limit. So, when we go from a stationary state to a non-stationary state, something very non-trivial happens. What exactly happens and whether this transition is continuous or not on a critical double scaling limit remains open and needs further investigation.

In a previous work, two of the authors have already shown numerically ~\cite{BK} that the universality class seems to be very large, as it extends to a gas of particles coupled by a collective term, which provides a weak coupling between particles~\cite{BM}, or to a gas of particles whose motion is driven 
by multiplicative stochastic processes or Kesten processes~\cite{K,B}. In \cite{BK}, we also found numerically that the Ornstein-Uhlenbeck process quite likely follows the same statistics, though strong finite-size effects prevented us at that time from obtaining a definite answer, which we have established in this work.

Further questions that remain open are, for example: how general can the potential $V(x)$ be?
Can it go beyond polynomial growth, and if so, at what growth rate will universality break down? We indeed think that at some point we may lose this universality once the potential exhibits a hard wall at its upper edge. Generally,
it would be interesting to work out a classification 
of universality classes, like that known for static order statistics \cite{FT,Gnedenko}, but for stochastic processes driven by random changes.
One may ask what the general mechanism is that creates this or that universality. 
What happens if one goes beyond Brownian motion in a potential and considers, for example, L\'evy flights? What would then be the universality classes? It might be intriguing to identify those classes and derive similar results for the overlap statistics.

It would be interesting to extend the methods discussed in this paper, 
which consider the system state at only two points in time (the beginning and end of a time interval), 
to the case of tracking the system state over the entire interval. This would open up the possibility 
of answering questions such as: what is the probability that a leader will consistently be the leader 
over the entire time interval? \cite{AR}.

Another interesting issue closely related to ranking, 
but going beyond order statistics, is the evolution of
the rank correlation coefficients \cite{Spearman,Kendall} for 
the whole population of particles undergoing Brownian motion.
Can one say how the rank correlation coefficient between 
ranks in the initial population and the population after time $t$ 
depends on $t$?

Finally, let us comment on the scaling relation that maps the original coordinates to local coordinates augmented for leader position analysis
\eqref{eq:scaling.position}. This is the standard scaling relation used in extreme value statistics. The map between the original and local coordinates is usually linear. This is particularly true for the uncorrelated systems discussed in this paper. However, for strongly correlated systems, nonlinear maps may be necessary. An example is the eigenvalue statistics of some random matrices at a spectral edge; see \cite{ABK,AGK}, where at the critical point of the phase transitions discussed there, the map to local coordinates had to be nonlinear. In other words, a linear map cannot always correctly capture the properties of the local spectral statistics near the spectral edges, where the largest and smallest eigenvalues are located. Of course, in this case, there is no reshuffling of the eigenvalue order due to eigenvalue repulsion. However, one can speculate that by slightly relaxing the interactions between eigenvalues, one can switch from Wigner statistics to Poisson statistics; see e.g. \cite{Johansson}, and allow the order of eigenvalues to be shuffled without changing the local statistics at the spectral edge.

\begin{acknowledgments}
The authors are grateful for inspiring discussions with Jean-Philippe Bouchaud, Pierre Le Doussal, and Satya Majumdar. ZB and TM thank the Polish Ministry
of Science and Higher Education for the financial support provided through a subsidy. MK is funded by the Australian Research Council through the Discovery Project Grant No. DP250102552.
\end{acknowledgments}


\begin{thebibliography}{99}

\bibitem{ranking} 
M.~Alvo and P.~L.~H.~Yu, Statistical Methods for Ranking Data, Springer, Heidelberg, 2014.

\bibitem{os1} 
J.~Galambos, The Asymptotic Theory of Extreme Order Statistics, R.E. Krieger Publishing. Co., Malabar, Florida, 1987.

\bibitem{os2} 
H.~N.~Nagaraja and H.~A.~David, Order statistics, (third ed.), Wiley, New Jersey, 2003.

\bibitem{os3} 
S.~N.~Majumdar and G.~Schehr, Statistics of Extremes and Records in Random Sequences (Oxford Graduate Texts), 2024. 

\bibitem{MPS} 
S.~N.~Majumdar, A.~Pal, and G.~Schehr, Physics Reports {\bf 840}, 1--32 (2020) [arXiv:1910.10667].

\bibitem{KLD}
D.~Brockington and J.~Warren, arXiv:2208.11952 (2022).

\bibitem{HCCC}
J.~B.~Hass, A.~N.~Carroll-Godfrey, E.~I.~Corwin, and I.~Z.~Corwin,  	Phys. Rev. E {\bf 107}, L022101 (2023) [arXiv:2205.02265].

\bibitem{HCC}
J.~B.~Hass, I.~Corwin, and E.~I.~Corwin, Phys. Rev. E {\bf 109}, 054101 (2024) [arXiv:2308.01267].

\bibitem{DDP}
S.~Das, H.~Drillick, and S.~Parekh,  	Journal of Functional Analysis {\bf 287}, 110609 (2024) [arXiv:2304.14279].

\bibitem{LX}
B.~Landon and T.~Xian, arXiv:2509.14192 (2025).

\bibitem{O} 
H.~A.~Orr, Genetics {\bf 163}, 1519 (2003). 

\bibitem{JK} 
K.~Jain and J.~Krug, J. Stat. Mech., P04008 (2005) [arXiv:q-bio/0501028].

\bibitem{BsM} 
I.~Bena and S.~N.~Majumdar, Phys. Rev. {\bf E}, 75, 051103 (2007) [arXiv:cond-mat/0701130].

\bibitem{JRBO} 
P.~Joyce, D.~R.~Rokyta,  C.~J.~Beisel, and H.~A.~Orr, Genetics {\bf 180}, 1627 (2008).

\bibitem{AMS} 
D.~Ben-Avraham, S.~N.~Majumdar, and S.~Redner,  J. Stat. Mech. {\bf2007}, L04002 (2007) [arXiv:physics/0702168].

\bibitem{LeD} 
P.~Le~Doussal, Phys. Rev. E  {\bf 109}, 024101 (2024) [arxiv:2308.16709].

\bibitem{MS}  
S.~N.~Majumdar and G.~Schehr, Phys. Rev. E {\bf 110}, 044111 (2024) [arXiv:2403.06964].

\bibitem{BK} 
Z.~Burda and M.~Kieburg, Phys. Rev. E {\bf 112}, 014114 (2025) [arXiv:2412.20818].

\bibitem{BGFSBBB} 
N.~Blumm, G.~Ghoshal, Z.~Forr\'o, M.~Schich, G.~Bianconi, J.-P.~Bouchaud, and A.-L.~Barab\'asi, Phys. Rev. Lett {\bf 109}, 128701 (2012).

\bibitem{BKMS} 
Z.~Burda, M.~J.~Krawczyk, K.~Malarz, and M.~Snarska, Entropy {\bf 23}, 842 (2021) [arXiv:2105.08048].

\bibitem{IPGB} 
G.~I\~niguez, C.~Pineda, C.~Gershenson, and A.-L.~Barab\'asi, Nature Communications {\bf 13}, 1646 (2022) [arXiv:2104.13439].

\bibitem{WK} 
M.~Wo\l oszyn and K.~Ku\l akowski, Physica A {\bf 610}, 128402  (2023) [arXiv:2210.10484].

\bibitem{DCLMN} 
F.~De~Domenico, F.~Caccioli, G.~Livan, G.~Montagna, and O.~Nicrosini, R. Soc. Open Sci. {\bf 11}, 240177 (2024).

\bibitem{KM} 
M.~Krawczyk and K.~Malarz, Chaos {\bf 34}, 073122 (2024).

\bibitem{DHJX} 
P.~Dong, R.~Han, B.~Jiang, and Y.~Xu, Journal of the Royal Statistical Society Series B: Statistical Methodology {\bf 88}, 221–238 (2026) [arXiv:2406.16507].

\bibitem{DDMS2017} 
D.~S. Dean, Pi.~Le~Doussal, S.~N.~Majumdar, and G.~Schehr, J. Stat. Mech. {\bf 2017}, 063301 (2017) [arXiv:1612.03954].

\bibitem{EM2008} 
M.~R.~Evans and S.~N.~Majumdar, J. Stat. Mech. {\bf 2008}, P05004 (2008) [arXiv:0804.0197].

\bibitem{stat-overlap}
B. Kjos-Hanssen, J. Log. Comput. {\bf 32}, 1611 (2022) [arXiv:2111.02498].

\bibitem{AW} 
J.~Abate and W.~Whitt, Advances in Applied Probability, {\bf 19}, 560 (1987).

\bibitem{R} 
H.~Risken, The Fokker–Planck Equation: Methods of Solution and 
Applications, 2nd Edition, Springer, 1996.

\bibitem{Gumbel}
E.~J.~Gumbel, Statistics of Extremes, Dover, New York, 1958.

\bibitem{BM} 
J.-P.~Bouchaud and M.~M\'ezard,  Physica A {\bf 282}, 536 (2000), [arXiv:cond-mat/0002374].

\bibitem{K} 
H.~Kesten, 
Acta Math. {\bf 131}, 207 (1973).

\bibitem{B} 
D.~Buraczewski, E. Damek, and T. Mikosch,  
Stochastic models with power-law tails, Springer, Heidelberg, 2016.

\bibitem{FT} 
R.~A.~Fisher and L.~H.~C.~Tippett, Mathematical Proceedings of the Cambridge Philosophical Society, {\bf 24}, 180 (1928).

\bibitem{Gnedenko}
B.~V.~Gnedenko, Ann. Math. {\bf 44}, 423 (1943).

\bibitem{SM}
G. Schehr and S.N. Majumdar, {\em Exact record and order statistics of random walks via first-passage ideas, a book chapter
in ”First-Passage Phenomena and Their Applications”}, Eds. R. Metzler, G. Oshanin, S. Redner, World Scientific (2013),
[arXiv: 1305:0639].

\bibitem{AR} We would like to thank an anonymous reviewer for drawing attention to this interesting class of problems.

\bibitem{Spearman}
C.~Spearman,  The American Journal of Psychology  {\bf 15}, 72 (1904).

\bibitem{Kendall}
 M.~G.~Kendall, Biometrika {\bf 30}, 81 (1938).

\bibitem{ABK}
 G.~Akemann, Z.~Burda, and M.~Kieburg, Phys. Rev. E {\bf 102}, 052134 (2020) [arXiv:2008.11470].

\bibitem{AGK}
 G.~Akemann, V.~Gorski, and M.~Kieburg, J. Phys. A {\bf 55}, 194002 (2022) [arXiv:2112.12447].

\bibitem{Johansson}
 K.~Johansson, Probab. Theory Relat. Fields {\bf 138}, 75 (2007) [arXiv:math/0510181].

\end{thebibliography}
\end{document}